\documentclass[oneside,english]{elsart}
\usepackage[T1]{fontenc}
\usepackage[latin1]{inputenc}
\usepackage{color}
\usepackage{graphicx}
\IfFileExists{url.sty}{\usepackage{url}}
                      {\newcommand{\url}{\texttt}}

\makeatletter

\newcommand{\lyxdot}{.}

\usepackage{babel}
\makeatother
\begin{document}
\begin{frontmatter}

\title{Measurement and Simulation of the Neutron Response of the Nordball
Liquid Scintillator Array}

\author[gla]{A. Reiter\thanksref{AR}},
\author[lu]{J.-O. Adler},
\author[isp]{I. Akkurt},
\author[gla]{J.R.M. Annand\thanksref{JRMA}},
\author[to]{F. Fasolo},
\author[lu]{K. Fissum},
\author[lu]{K. Hansen},
\author[lu]{L. Isaksson},
\author[lu]{M. Karlsson},
\author[lu]{M. Lundin},
\author[gla]{J.C. McGeorge},
\author[lu]{B. Nilsson},
\author[gla]{G. Rosner},
\author[lu]{B. Schr\o der} and
\author[to]{A. Zanini}

\address[gla]{Department of Physics \& Astronomy, University of Glasgow, Glasgow
G12 8QQ, Scotland, UK.}

\address[lu]{Department of Physics, University of Lund, S\"olvegatan 14, S-223
62, Lund, Sweden.}

\address[to]{Istituto Nazionale di Fisica Nucleare (INFN), Sez. Torino, Via
P. Giuria 1, 10128 Torino, Italy.}

\address[isp]{Süleyman Demirel University, Fen-Edebiyat Faculty, 32 260 Isparta,
Turkey.}

\thanks[AR]{Present Address: Gesellschaft für Schwerionenforschung mbH, Planckstrasse
1, 64291 Darmstadt, Germany.}

\thanks[JRMA]{Corresponding author: j.annand@physics.gla.ac.uk, +44 141 330
6428.}

\begin{abstract}
The response of the liquid scintillator array Nordball to neutrons
in the energy range $1.5<T_{n}<10$~MeV has been measured by time
of flight using a $^{252}$Cf fission source. Fission fragments were
detected by means of a thin-film plastic scintillator. The measured
differential and integral neutron detection efficiencies agree well
with predictions of a Monte Carlo simulation of the detector which
models geometry accurately and incorporates the measured, non-linear
proton light output as a function of energy. The ability of the model
to provide systematic corrections to photoneutron cross sections,
measured by Nordball at low energy, is tested in a measurement of
the two-body deuteron photodisintegration cross section in the range
$\mathrm{E}_{\gamma}=14-18$~MeV. After correction the present $^{2}\mathrm{H\left(\gamma,n\right)p}$
measurements agree well with a published evaluation of the large body
of $^{2}\mathrm{H\left(\gamma,p\right)n}$ data. 
\end{abstract}
\end{frontmatter}

\section{\label{sec:Introduction}Introduction}

Recently photoneutron production cross sections on a range of nuclei
\cite{Akkurt} have been measured in the energy region 11-30~MeV
at the high-duty-factor, tagged-photon facility of MAX-lab \cite{JOA1}
in Lund, Sweden. The purpose is to test calculations of neutron dose
received during the course of bremsstrahlung radiotherapy \cite{Zanini}.
In these time-of-flight (TOF) experiments, at relatively low kinetic
energy ($T_{n}$), the neutron signal is obscured by an accelerator-induced
room background and a high rate of random-coincidence events produced
by untagged photons incident on the experimental target. The bulk
of this background arises from detected photons, so that good particle
identification is required to access $T_{n}\sim1$~MeV, where both
the photoproduction cross section and the biological effectiveness
of neutrons are at a maximum. The MAX-lab liquid scintillator array
Nordball \cite{Nordball}, with good $n/\gamma$ pulse shape discrimination
(PSD) properties and $\sim1$~ns (FWHM) time resolution, was used
for these measurements. This paper presents Nordball calibration procedures
and a computer model used to evaluate systematic effects which distort
the measured neutron yield. 

At low energies neutron attenuation and multiple scattering effects
on measured neutron yields are large. Thus a simulation of the experiment
based on GEANT-3 \cite{Geant3}, which models neutron interactions
in all materials in the vicinity of the detector array, has been developed
to correct for these effects. The non-linear pulse-height response
of the scintillators to low energy recoil protons has a critical bearing
on these calculations, especially close to threshold, and so this
was measured with a white $^{252}$Cf neutron source. This is a calibration
standard since the neutron yield from spontaneous fission (decay branching
ratio 3.1\%) is high and both the absolute numbers and energy spectrum
of prompt neutrons per fission are well known \cite{Bowman,Mannhart}.
Neutron detector response may be obtained in a TOF measurement if
fission events are tagged by detection of at least one heavy fragment,
which provides a time reference. Although fission chambers \cite{Cub}
are possibly the optimal detectors for fragment detection, a simple
thin-film organic scintillator presented a viable, available alternative
at the time of the measurement. This type of detector has several
desirable properties:

\begin{enumerate}
\item Insensitivity to $\gamma$-rays or neutrons as the plastic is extremely
thin.
\item Fast response with a similar time resolution to Nordball, which also
gives high counting rate capability.
\item Fast, cheap production. 
\end{enumerate}
Here we compare the measured and simulated fission-neutron response
of the Nordball array and describe a test of the simulation by measurement
of the well known $^{2}\mathrm{H\left(\gamma,n\right)p}$ cross section.
Here the simulations have been used to compute the large attenuation
and multiple scattering corrections, as well as the neutron detection
efficiency. Sec. \ref{sec:Time-of-flight-experiment} gives an overview
of the experimental setup, the measured Nordball response is presented
in Sec. \ref{sec:Nordball-Performance}, the Monte Carlo (MC) simulation
is described in Sec. \ref{sec:The-GEANT-3-based}, measurements and
simulations are compared in Sec. \ref{sec:Detection-efficiency},
and a short summary is given in Sec. \ref{sec:Conclusions}.

\section{\label{sec:Time-of-flight-experiment}The time-of-flight experiment}

\subsection{\label{sub:Nordball-array}The Nordball array}

\begin{figure}[th]
\includegraphics[width=1\columnwidth,keepaspectratio]{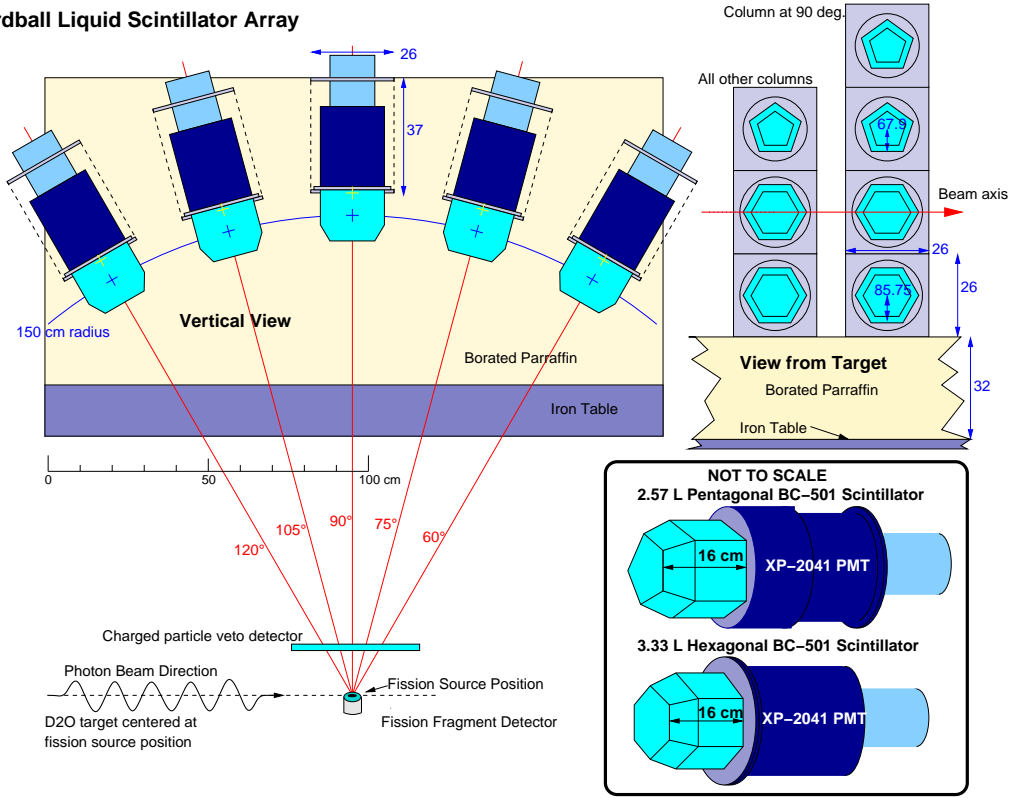}

\caption{\label{cap:Nordball-detector-geometry}Layout of Nordball array for
$^{252}$Cf fission-source and $^{2}\mathrm{H\left(\gamma,n\right)p}$
measurements.}
\end{figure}

The Nordball detector (Fig. \ref{cap:Nordball-detector-geometry})
consists of 16 liquid scintillators, type Bicron BC-501, with PSD
capability. Ten detectors are of hexagonal cross section and six detectors
of pentagonal cross section. Their respective volumes are 3.3~l and
2.6~l at a common thickness of 16~cm. The liquid is contained within
a 2~mm thick, stainless steel cannister, coated on its inner surface
with TiO$_{2}$ reflective paint. This is connected to a 5~in. XP2041
photomultiplier tube (PMT) via a Pyrex glass window. A cylinder of
$\mu$-metal shields the PMT from stray magnetic fields and an outer
plastic housing encases tube and voltage-divider circuit, which provides
a negative anode signal.

All detectors were mounted on aluminium frames and placed on a 32~cm
thick layer of borated paraffin, supported by an iron table. The configuration
of Fig. \ref{cap:Nordball-detector-geometry} consisted of five detector
columns, positioned in 15$^{\circ}$ steps, at a distance of 150~cm
from the central position where \textcolor{red}{}\textcolor{black}{the
experimental target or} the fission detector was located. The bigger
hexagonal detectors were placed in the two bottom rows, five pentagonal
detectors in the third, and the last one on top of the central column.
Paraffin towers to both sides of the iron table partially shielded
the array against regions of strong neutron background. For the deuteron
photodisintegration experiment a $40\times40\times1.8$~cm plastic
scintillator sheet was inserted between the target and Nordball to
identify charged particle events. 

\subsection{\label{sub:Fission-detector}Fission detector}

The thin-film scintillator was prepared following Ref. \cite{Ajitanand}.
A solution of plastic scintillator in xylene was spread uniformly
over the glass window of a 2~in. PMT (\textcolor{black}{Philips}
XP2262B) which was placed upright in a vacuum chamber. Slow evacuation
of the chamber causes solvents to evaporate and a thin layer of scintillating
material remains on the glass. Its thickness was tuned, by varying
the amount of plastic in the solution, to optimise discrimination
between fission fragments and the 30 times more numerous $\alpha$
particles. 

The fission source \cite{Karlsson} consists of a platinum-clad, nickel
disk, on to which $\mathrm{Cf_{2}O_{3}}$ was electro-deposited. The
active area (0.2~cm$^{2}$) is covered by a 50~$\mu$g/cm$^{2}$
layer of gold allowing the passage of fission fragments with relatively
small energy loss. The source was placed on the centre of the scintillating
film and the PMT was sealed with a plastic cap, lined with aluminised
mylar foil, to exclude external light and provide some reflection
of scintillation light. The fission detector was positioned so that
the source sat at the target-centre position (Fig. \ref{cap:Nordball-detector-geometry})
at the same height as the middle of the second Nordball row and the
photon beam axis. 

The axis of the PMT was offset $\sim10^{\circ}$ from vertical. \textcolor{black}{Apart
from the bottom row of the array, this offset avoided fission neutrons
having to pass through the glass of the PMT on the way to the Nordball
detectors. Placing the PMT horizontally would have displaced the loose
source from the active area of the PMT face. As a consequence of attenuation
in glass, the detected neutron yield in the bottom detectors was found
to be about 15\% smaller at low energies where the interaction cross
section is highest. These losses are consistent with the predictions
of a computer model of the experiment (Sec. \ref{sub:Geant3-Stanton-code})
which approximated the PMT by a 2~mm thick glass cylinder and calculated
for tilting angles in the range of 5--15 degrees. However the angle
was not sufficiently well determined to enable quantitative comparisons}
and hence the bottom detectors were excluded from the analysis of
differential and absolute neutron detection efficiencies presented
in Sec. \ref{sec:Detection-efficiency}.

No suppression of bottom-row neutron yield was observed in the deuteron
photodisintegration measurement, which used a cylindrical $\mathrm{D_{2}O}$
target, supporting the assumption of increased neutron attenuation
in glass during the $^{252}$Cf measurement. 

\begin{figure}[th]
\includegraphics[width=1\columnwidth,keepaspectratio]{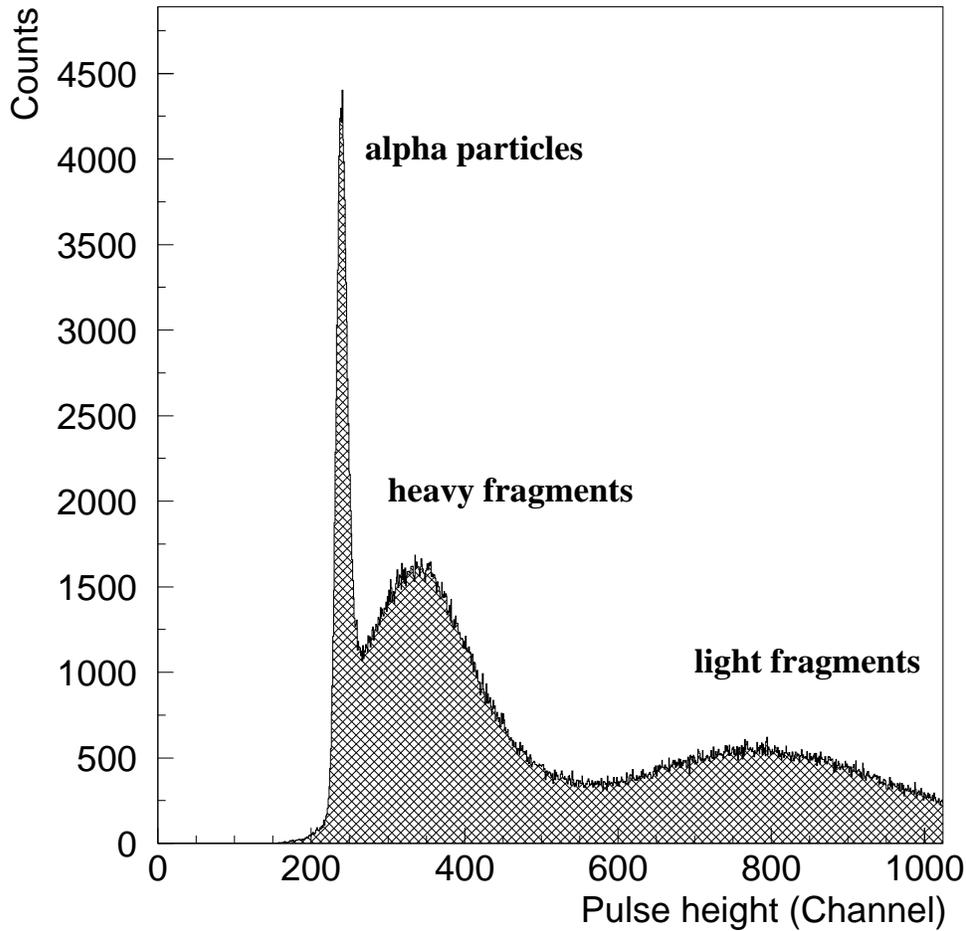}

\caption{\label{cap:Measured-pulse-height}$\mathrm{^{252}Cf}$ pulse-height
spectrum obtained with the thin-film fission detector. The narrow
peak results from those alpha particles remaining above pulse-height
threshold. The two broad distributions correspond to heavy and light
fission fragments.}
\end{figure}

Fig. \ref{cap:Measured-pulse-height} shows the fragment pulse-height
distribution obtained from the $^{252}$Cf source, which produced
a counting rate of $\sim3$~kHz. With $\sim2\pi$ acceptance for
charged particle detection, the effective thickness of the scintillator
depended on emission angle and not all 6.11~MeV $\alpha$ particles
could be separated from heavy fragments by pulse height. A small fraction
of $\alpha$ particles had to be accepted in the trigger, resulting
in the sharp spike observed around channel 250. These events are completely
uncorrelated with any detected neutron, confirming that they are not
produced by fission fragments.  The two broad distributions represent
interactions of heavy and light fission fragments, \textcolor{black}{with}
\textcolor{black}{$\sim10$\% of the former lost, judging from the
width of the peak.} An estimate of the resulting distortion of the
neutron spectrum was based on Ref. \cite{Bowman} where the shapes
of the neutron energy distributions and the mean numbers of emitted
neutrons per fission event for light (1.97) and heavy (1.70) fragments
are presented. The distortion in the neutron spectrum is below 2\%,
and the mean number of neutrons $\overline{\nu}$ is reduced by 5.4\%.
No significant deviations from this simple estimate were observed
when measured neutron spectra, corresponding to different regions
in the fission pulse-height distribution, were compared. However rather
than attempt a more sophisticated correction we have factored these
distortion effects into the systematic uncertainties associated with
the evaluation of the $^{252}$Cf fission neutron yield \cite{Mannhart}.

\subsection{\label{sub:Electronics}Electronics}

\begin{figure}[th]
\includegraphics[width=1\columnwidth,keepaspectratio]{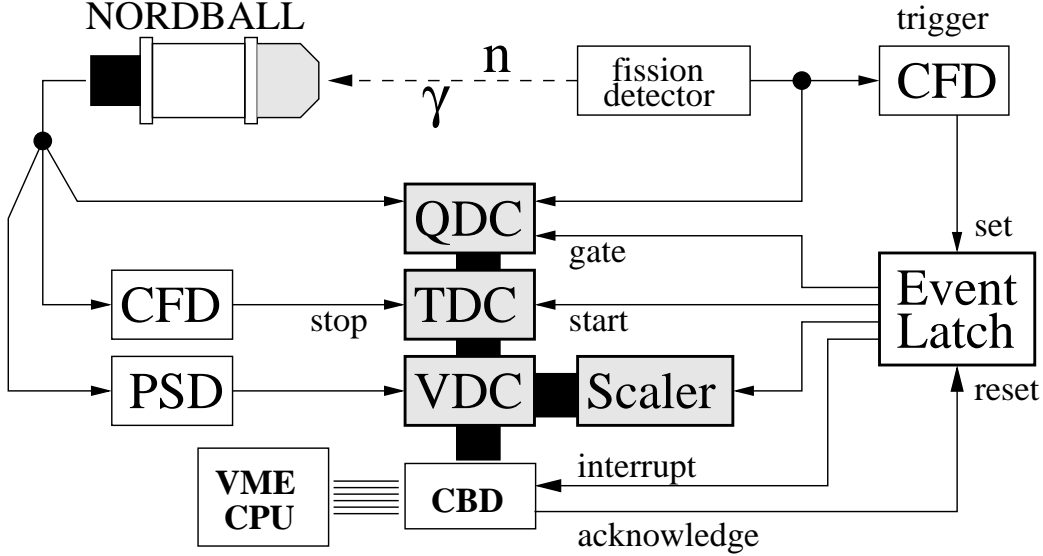}

\caption{\label{cap:Block-diagram-of}Block diagram of the experimental electronics
(Sec. \ref{sub:Electronics}).}
\end{figure}

The layout of the electronics is shown in Fig. \ref{cap:Block-diagram-of}
and the data acquisition and analysis system ACQU is described in
Ref. \cite{jrma1}. Analogue to digital converters (ADC) and scalers
were read out by a VMEbus single board computer via the CAMAC bus
and a VMEbus CAMAC parallel branch driver (CBD). During readout the
event latch prevented further triggers until reset at the end of the
event.

The experimental trigger for the $^{252}$Cf measurement was made
by the constant-fraction discriminator (CFD) of the fission detector.
The number of triggers was recorded by a scaler for normalisation
purposes and the fission pulse height was recorded in a charge-integrating
ADC (QDC). Gate and start signals for the various ADCs were derived
from the trigger. The anode signal of each Nordball scintillator was
split 3 ways and fed to:

\begin{enumerate}
\item a QDC to record the pulse amplitude.
\item a CFD connected to a time to digital converter (TDC) to record the
time of flight to the scintillator.
\item specialist hardware developed for fast pulse-shape analysis \cite{jrma2},
which produces a {}``pulse-shape'' output, recorded in a voltage
to digital converter (VDC).
\end{enumerate}
The electronics for the $^{2}\mathrm{H\left(\gamma,n\right)p}$ measurement
were very similar, except that the trigger was made by the logical
OR of the Nordball CFD outputs. In this case the logic outputs from
the pulse-shape analysers, flagging a neutron, were used to gate the
interrupt signal to the CPU. In addition the signals from each of
the 64 focal-plane detectors of the photon-tagging spectrometer were
recorded in TDCs and scalers.

\section{\label{sec:Nordball-Performance}Nordball Performance}

\subsection{\label{sub:Pulse-height-calibration}Pulse-height calibration}

The light output from the Nordball detectors was calibrated with the
$\gamma$-ray sources $^{60}\mathrm{Co}$, $^{228}\mathrm{Th}$, and
Pu/Be. Following Ref. \cite{Knox}, the edges of the resulting Compton
electron spectra were initially assumed to be the channels containing
90\% of the counts in the peaks in the Compton end-point pulse-height
distributions. However this procedure may be over simplistic as the
Compton edge position depends on the resolution of the detection system
under consideration. In Ref. \cite{Dietze} an empirical formula was
determined to estimate the resolution from measured pulse-height spectra
and tables of resolution-dependent Compton edge positions, relative
to the channel of maximum counts, were produced. These tables were
checked using the present GEANT-3 simulation (Sec. \ref{sec:The-GEANT-3-based}),
which reproduced the result for both Nordball and the detector geometries
of Ref. \cite{Dietze}. \textcolor{black}{However the simple approximation
of Ref. \cite{Knox} does not deviate from Ref. \cite{Dietze} by
more than 3\%.}

The pulse-height resolution (FWHM) of the Nordball array was parameterised
\cite{Dietze}, in terms of scintillation light output $L$ in units
of electron energy equivalent (MeVee), as:

\begin{equation}
\frac{\Delta L}{L}=\sqrt{\alpha^{2}+\beta^{2}/L+\gamma^{2}/L^{2}}\label{eq:light-output}\end{equation}

Small non-linearities in the low energy electron response \cite{Schoelermann}
were neglected. Since the individual detector responses were very
similar, they were summed before least-squares fitting to obtain the
average response of the Nordball array. Parameter values of $\alpha=3.2$
$\pm$10 \% and $\beta=10.9$ $\pm$5 \% were obtained with $\gamma$
fixed at zero and this resolution function was implemented in the
MC simulation. The obtained values are similar to those of Ref. \cite{Filchenkov,Buermann,Naqvi}.

\subsection{\label{sub:Particle-identification}Particle identification}

\begin{figure}[th]
\includegraphics[width=1\columnwidth,keepaspectratio]{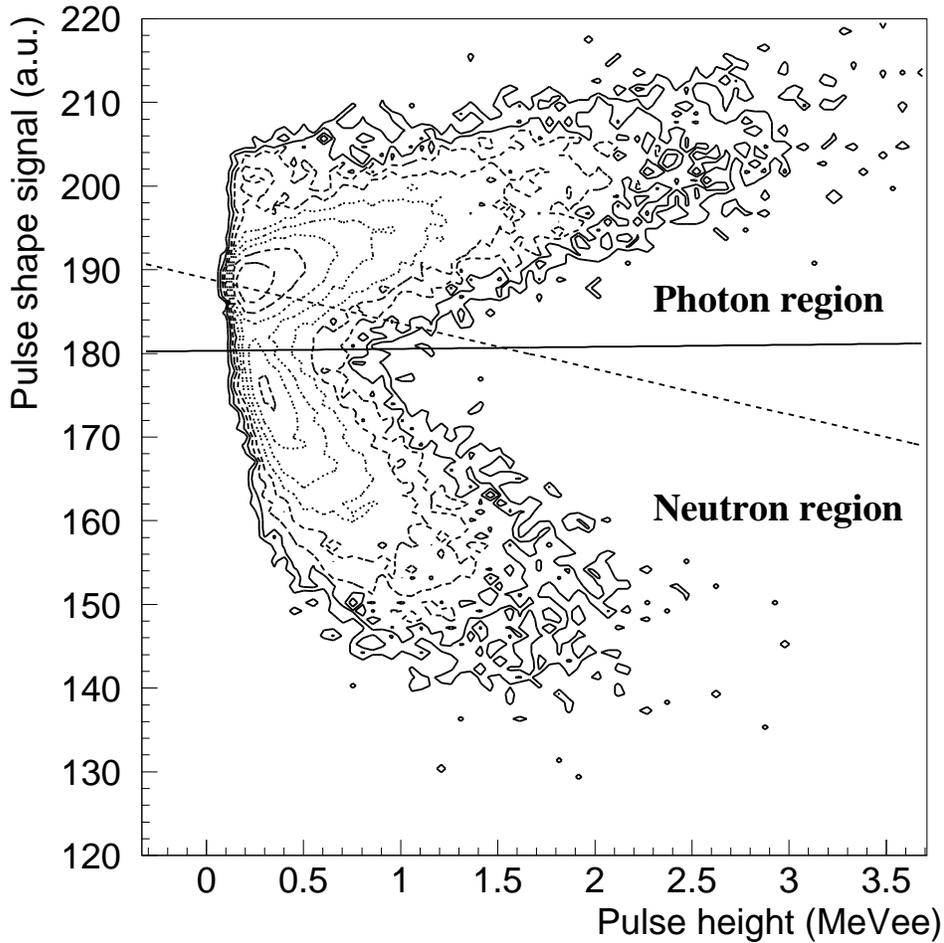}

\caption{\label{cap:Neutron-photon-discrimination-plot}Neutron-photon discrimination
plot for a single Nordball detector, showing the pulse-shape information
as a function of pulse height. The dashed/full lines represent the
boundaries of the neutron/photon regions.}
\end{figure}

Pulse shape discrimination (Fig. \ref{cap:Neutron-photon-discrimination-plot})
was applied to separate detected neutrons and photons. To quantify
neutron and photon regions, two-Gaussian fits were made to pulse shape
projections generated for 50 keVee-wide pulse-height bins, yielding
peak positions and widths for each distribution. Intervals of $\pm3\sigma$
around the peak positions defined the neutron or photon region limits
at a given pulse height and straight-line fits to these limits defined
the boundaries of the neutron/photon regions of the pulse shape plot.
Neutron events below the dashed line of Fig. \ref{cap:Neutron-photon-discrimination-plot}
have some photon contamination at pulse heights below $\sim1$~MeVee.

\subsection{\label{sub:Response-function}The neutron pulse-height response function}

\begin{figure}[th]
\includegraphics[width=1\columnwidth,keepaspectratio]{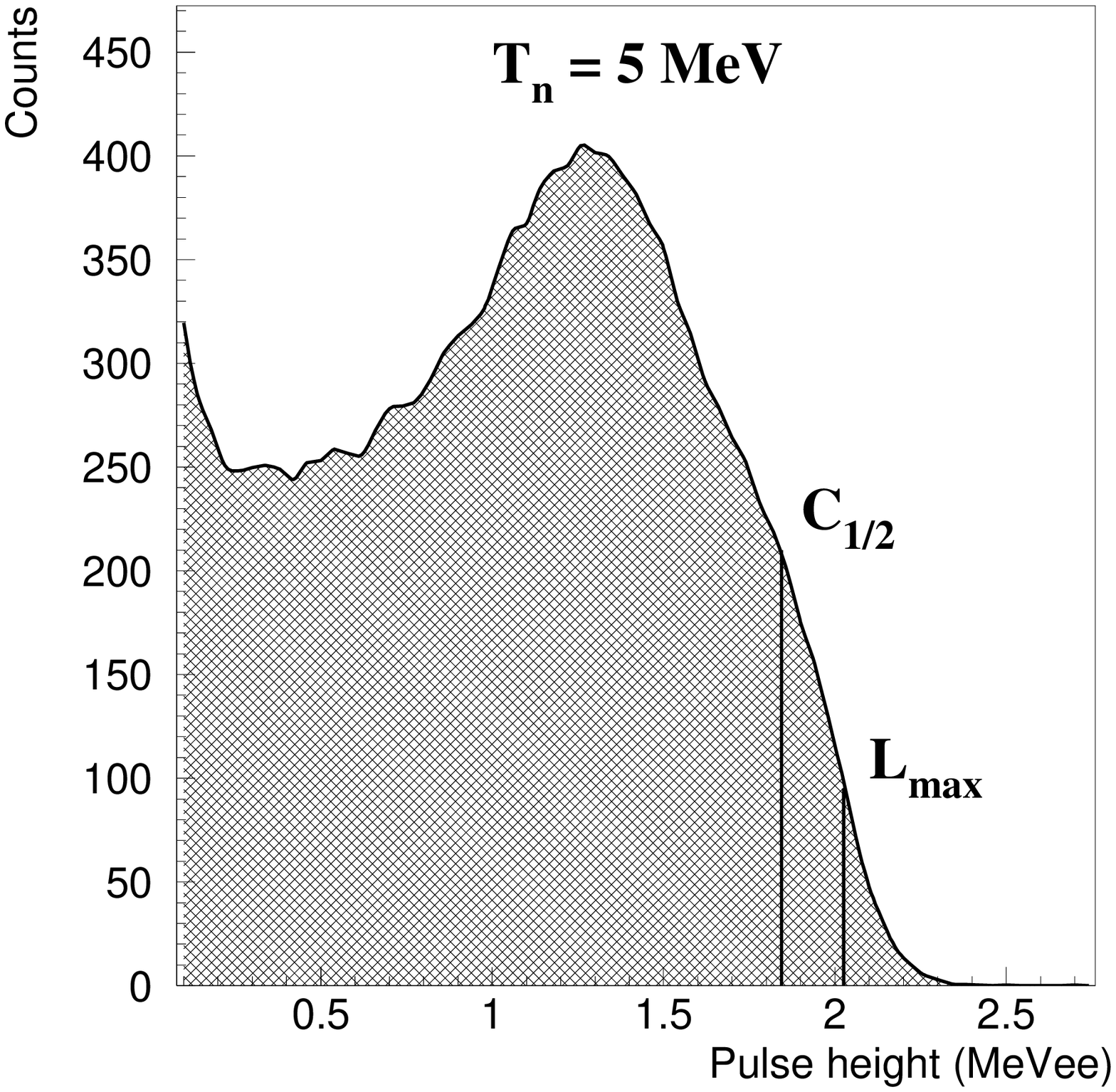}

\caption{\label{cap:corr_eff}\textcolor{black}{The simulated pulse-height
distribution for 5~MeV neutrons. }}
\end{figure}

The non-linear scintillation response to recoil protons is vital input
to a neutron efficiency simulation and must be measured to avoid large
systematic errors close to detection threshold. Using TOF information,
neutron events were divided into 625~keV-wide kinetic-energy bins
in the range $1.25<T_{n}<10$~MeV. The resulting pulse-height spectra
for each detector were summed to produce the overall response of the
array. For each value of $T_{n}$ the maximum proton recoil energy
was determined from the half-height position $C_{1/2}$($T_{n}$),
corrected by a multiplicative factor $C(T_{n})$ to account for multiple
neutron interactions in the scintillators. 

Numerical values for $C(T_{n})$ were derived from the MC simulation
(Sec. \ref{sub:Modified-Stanton}), which mimics the 625~keV neutron
energy bins used in the data analysis. \textcolor{black}{A pulse-height
distribution for 5 MeV neutrons is shown in Fig. \ref{cap:corr_eff}.
The half-height position $C_{1/2}$ at 1.85~MeVee underestimates
the expected light output of 2.03~MeVee for maximum-energy recoil
protons, calculated from the scintillator response function. This
leads to a correction factor $C(T_{n})=1.10$ at 5~MeV, while for
2~MeV neutrons $C(T_{n})=1.22$.} A 2nd-order polynomial fit to the
MC calculations in the range $T_{n}=1.25-10$~MeV yielded:

\begin{equation}
C(T_{n})=1.335-0.067\cdot T_{n}+0.004\cdot T_{n}^{2}\label{eq:C(T)}\end{equation}

\begin{figure}[th]
\includegraphics[width=1\columnwidth,keepaspectratio]{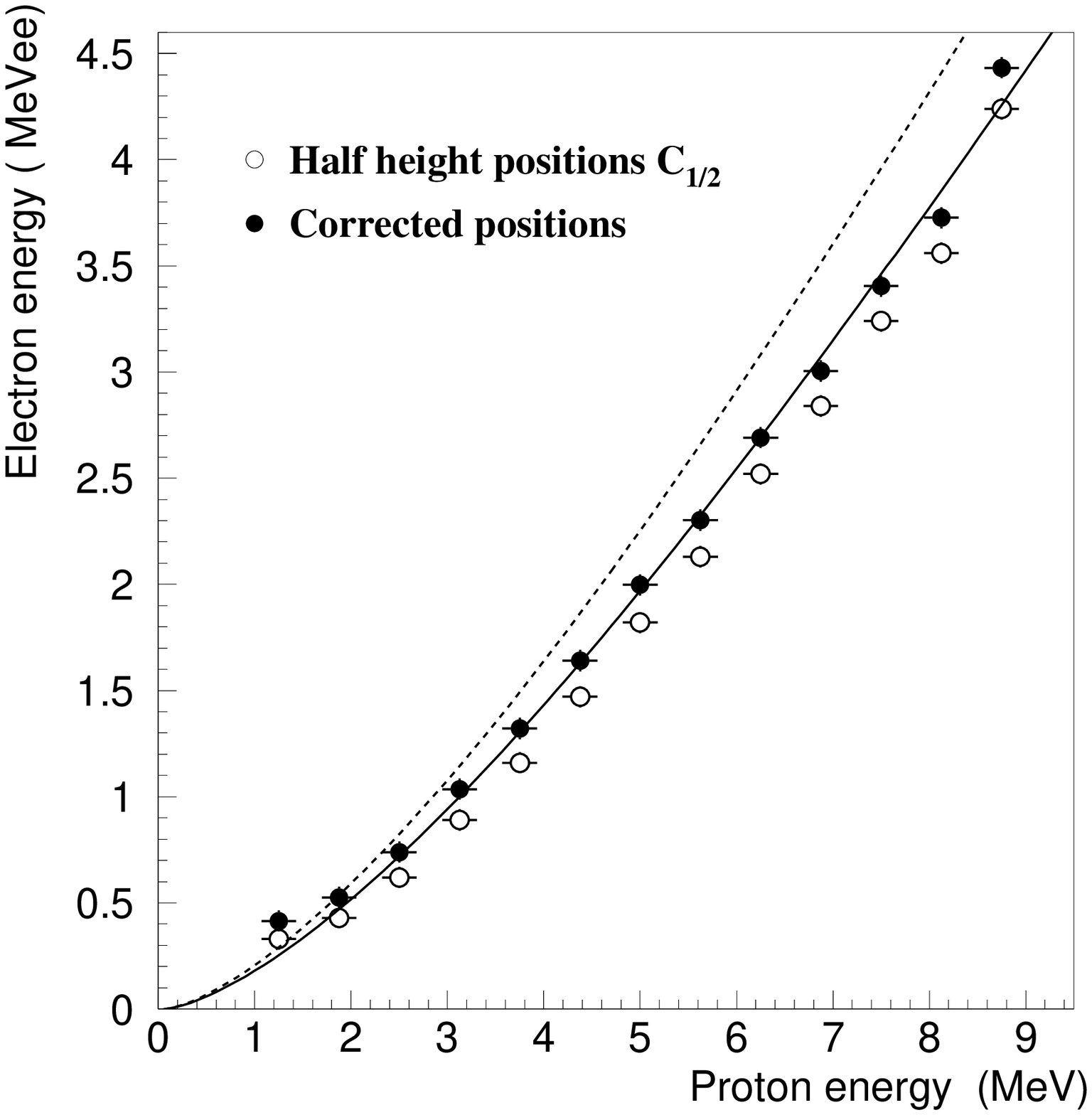}

\caption{\label{cap:Response-function-of}The pulse-height response function
of the Nordball array: full/empty circles represent the half-height
positions with/without correction for multiple interactions. The dashed
line represents the response function of Ref. \cite{Cecil}, the full
black line the same function multiplied by a factor of 0.9.}
\end{figure}

The proton response functions obtained with (filled circles) and without
(empty circles) the correction factors are shown in Fig. \ref{cap:Response-function-of}.
The symbol size is equivalent to the 50 keVee uncertainty of the half-height
position $C_{1/2}$, estimated from variations in the spline functions
which fitted the edge regions of the pulse-height spectra. The response
function of Ref. \cite{Cecil} (dashed line) describes the energy
dependence of the data quite well, after multiplication by a factor
of 0.9 (full line) which was determined from a least-squares fit.
Note that the first data point at 1.25 MeV is too high because the
hardware threshold of 0.25 MeVee cuts into the recoil proton edge
distribution (Sec. \ref{sub:Differential-detection-efficiency}).
The response of individual detectors showed no significant deviations
and a common light output function for the Nordball array was implemented
in the MC simulation. 

\section{\label{sec:The-GEANT-3-based}The GEANT-3 based Monte Carlo simulation}

Measurements of low energy neutron photoproduction require correction
for sizable systematic effects on the neutron acceptance, which has
led to the development of a GEANT-3 \cite{Geant3} based MC simulation,
into which neutron tracking methods from the STANTON code \cite{Stanton,Cecil}
have been incorporated. This coupled approach allows the use of the
GEANT-3 neutron cross section data base (package GCALOR \cite{Zeitnitz}
based on ENDF/B-6 \cite{ENDF}) at $T_{n}<$20~MeV and STANTON procedures
for $20<T_{n}<500$~MeV, where the latter provides a more detailed
and accurate representation of neutron interactions compared to standard
GEANT-3.

The flexible geometry framework of the MC simulation allowed comparison
of predictions with a broad range of published neutron response measurements.

\subsection{\label{sub:Modified-Stanton}The modified STANTON code}

A modified version of STANTON, augmented in the following ways, was
used to model the Nordball response in the $^{252}$Cf measurement:

\begin{enumerate}
\item The hexagonal Nordball geometry was added. Pentagonal elements were
treated as hexagonal, but scaled down in cross-sectional area to give
the correct pentagonal-element volume
\item \textcolor{black}{Isotropic neutron emission was added to the existing
event generator for the $^{252}$Cf source.}
\item The pulse-height resolution function of Sec. \ref{sub:Pulse-height-calibration}
was incorporated.
\item The scintillator response to proton interactions, as described in
Sec. \ref{sub:Response-function}, was incorporated. 
\item Results were stored as CERN HBOOK ntuples for later analysis.
\end{enumerate}
The core of the simulation, including the neutron tracking routines
of Ref. \cite{Cecil}, remained unchanged. After benchmark comparisons
which reproduced a variety of published neutron efficiency measurements
\cite{Drosg,Thornton,Al-Ohali}, the MC code was applied to Nordball
(Sec. \ref{sub:Differential-detection-efficiency},\ref{sub:Neutron-detection-efficiency}).

\subsection{\label{sub:Geant3-Stanton-code}The coupled GEANT3-STANTON code }

The geometry and materials of a particular experiment (Fig. \ref{cap:Nordball-detector-geometry})
can be described in fine detail using GEANT-3 but, designed primarily
for high energy physics, this fails to model low energy neutron interactions
accurately. However, with the addition of the GCALOR package the GEANT-3
description of $T_{n}<20$~MeV neutron interactions is quite accurate
due to the use of recent Evaluated Nuclear Data Files. These files
contain neutron partial cross sections, angular distributions and
secondary energy distributions evaluated on a grid of discrete energy
and angle points, for elements over the full range of the periodic
table. This is necessary as few-MeV neutron cross sections are large
and strongly dependent on energy and target mass. Any realistic model
requires consideration not only of neutron interactions within the
detector volume, but also of attenuation and multiple scattering effects
in all bulk materials in the locality of the detector. At higher energy
such effects become less important, but none the less it remains important
to model interactions in the detector volume as accurately as possible.

The present implementation of GEANT-3 incorporates STANTON methods
as a subroutine which tracks neutrons inside the organic-scintillator
detector volumes, while GCALOR is used for interactions outside the
detector volume. The geometry of all components considered in the
simulation, and hence calculation of volume boundaries, is controlled
entirely by the GEANT-3 tracking system. Original STANTON geometry
functions were replaced by the GEANT-3 subroutine \emph{gmedia,} which
may be tailored to handle pentagonal and hexagonal shaped detectors
in a straightforward way.

A test of the simulation is described in Sec. \ref{sub:Deuteron-photodisintegration},
where it has been used to calculate systematic corrections to a measurement
of the well known, two-body deuteron photodisintegration cross section.

\section{\label{sec:Detection-efficiency}The detection efficiency from the
$^{252}$Cf measurement}

\subsection{\label{sub:Differential-detection-efficiency}Differential detection
efficiency}

Normalised pulse-height spectra were sorted into incident neutron
energy bins of 0.5~MeV width and compared to the MC predictions.
A PSD cut was applied to select neutrons and events with unreasonably
large recoil proton pulse height, compared to that expected from the
incident neutron energy and measured proton response, were rejected.

\begin{figure}[th]
\includegraphics[width=1\columnwidth,keepaspectratio]{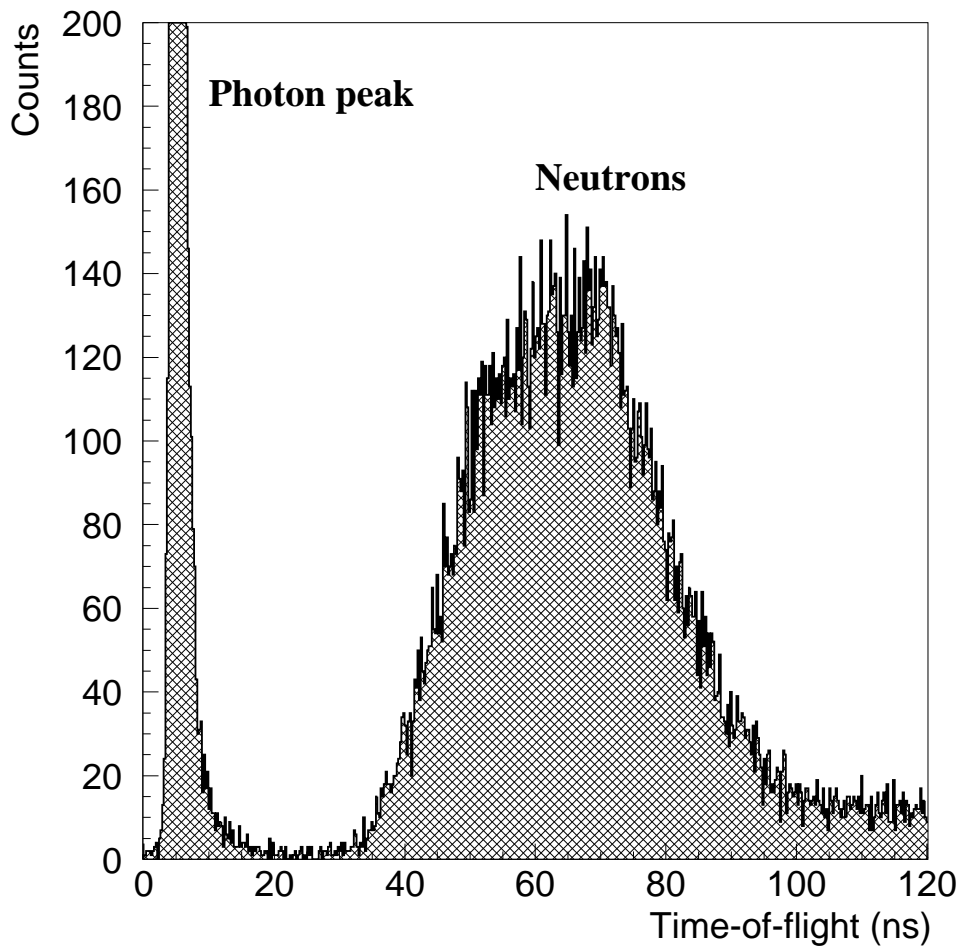}

\caption{\label{cap:Time-of-flight}Time of flight spectrum for a single Nordball
detector. The width of the photon peak is 1.7~ns (FWHM).}
\end{figure}

The resulting TOF spectrum for a hardware threshold of $\sim0.25$~MeVee,
shown in Fig. \ref{cap:Time-of-flight}, is virtually free of random
coincidences which would fill in the 20-30~ns TOF region between
the photon peak (some photons beat the PSD cut) and the neutron distribution.
A low energy tail due to neutrons that scattered en route to the detectors
(and hence had a longer flight path) is present at flight times of
100~ns and longer. These events were suppressed when the detection
threshold was raised, and above 0.7~MeVee they were virtually eliminated
as \textcolor{black}{discussed in Sec. \ref{sub:Particle-identification}.}

Assuming isotropic fission neutron emission and neglecting absorption
effects in air and detector canisters, which were estimated to be
below 3\%, the differential efficiency is given by: 

\begin{equation}
\frac{d\varepsilon_{exp}}{dL}(T_{n})=\frac{\frac{dN_{n}}{dL}(T_{n})}{\overline{\nu}\cdot N_{fiss}\cdot f_{n}(T_{n})\cdot\Delta T_{n}\cdot\Delta\Omega/4\pi}\label{eq:diffeff}\end{equation}

where $\frac{dN_{n}}{dL}(T_{n})$ is the pulse-height spectrum for
a neutron energy bin of width $\Delta T_{n}$ centered at $T_{n}$,
$\overline{\nu}=3.76$ (5.4\%) is the average number of prompt neutrons
per fission event, $N_{fiss}$ is the number of fission fragments
counted, $f_{n}(T_{n})$ (3.5\%) is the energy distribution of prompt
neutrons and $\Delta\Omega$ (0.4\%) is the solid angle subtended
by a detector. Relative uncertainties are given in the brackets. The
main sources of uncertainty result from distortions of $\overline{\nu}$
and $f_{n}(T_{n})$ due to not recording the full fission-fragment,
pulse-height spectrum (Sec. \ref{sub:Fission-detector}) and the $f_{n}(T_{n})$
uncertainty contains a further factor due to the uncertainty in $T_{n}$. 

\begin{figure}[th]
\includegraphics[width=1\columnwidth,keepaspectratio]{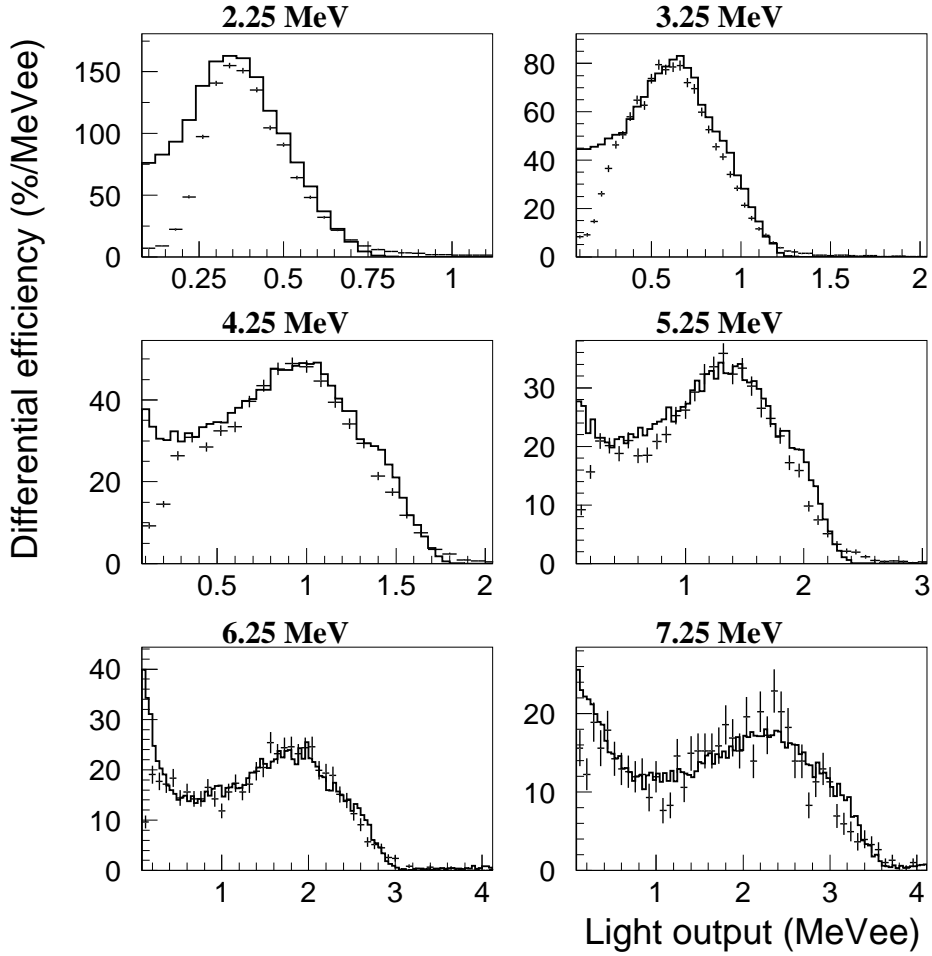}

\caption{\label{cap:Differential-detection-efficiency}Differential detection
efficiencies (Eq.\ref{eq:diffeff}) at neutron energies 2.25 - 7.25~MeV,
for 0.5~MeV-wide bins of neutron energy. The full-line histogram
is derived from the MC simulation, while the data points are derived
from the measured pulse-height response. Error bars denote statistical
uncertainties.}
\end{figure}

The agreement between data and simulation is generally good, as shown
in Fig. \ref{cap:Differential-detection-efficiency}. Since the statistical
uncertainty increases rapidly as neutron energy increases, the spectra
were binned a factor of 2 more coarsely above 5~MeV. The hardware
threshold cuts noticeably into the pulse-height distribution for values
below 0.3~MeVee. This (voltage) threshold is not sharp due to pulse
shape dependence and hence difficult to model accurately. Thus 0.3
MeVee represented a minimum \textcolor{red}{\emph{}}software threshold
which could be used for quantitative analysis. At low incident energies
multiple scattering leads to a more peak-like shape of the spectrum
which flattens out as the incident neutron energy increases. Above
6~MeV competing carbon cross sections lead to an enhancement of pulse
heights below $\sim0.5$~MeVee. 

\subsection{\label{sub:Neutron-detection-efficiency}Integral detection efficiency}

The neutron yield $N_{n}(T_{n})$ was obtained for detection thresholds
of 0.3, 0.5, 0.7, and 1.0~MeVee from the TOF distribution (Fig. \ref{cap:Time-of-flight}),
converted to $T_{n}$ assuming a path length of 1.5~m. Only events
with $T_{n}>$1~MeV (TOF below 110~ns) where a software pulse-height
threshold could be set above the uncertain hardware threshold range
were analysed. Substituting $N_{n}(T_{n})$ for $\frac{dN_{n}}{dL}\left(T_{n}\right)$
in Eq.\ref{eq:diffeff} delivered the absolute detection efficiency,
but the efficiency could have been calculated alternatively by integration
of the pulse-height spectra.

\begin{figure}[th]
\includegraphics[width=1\columnwidth,keepaspectratio]{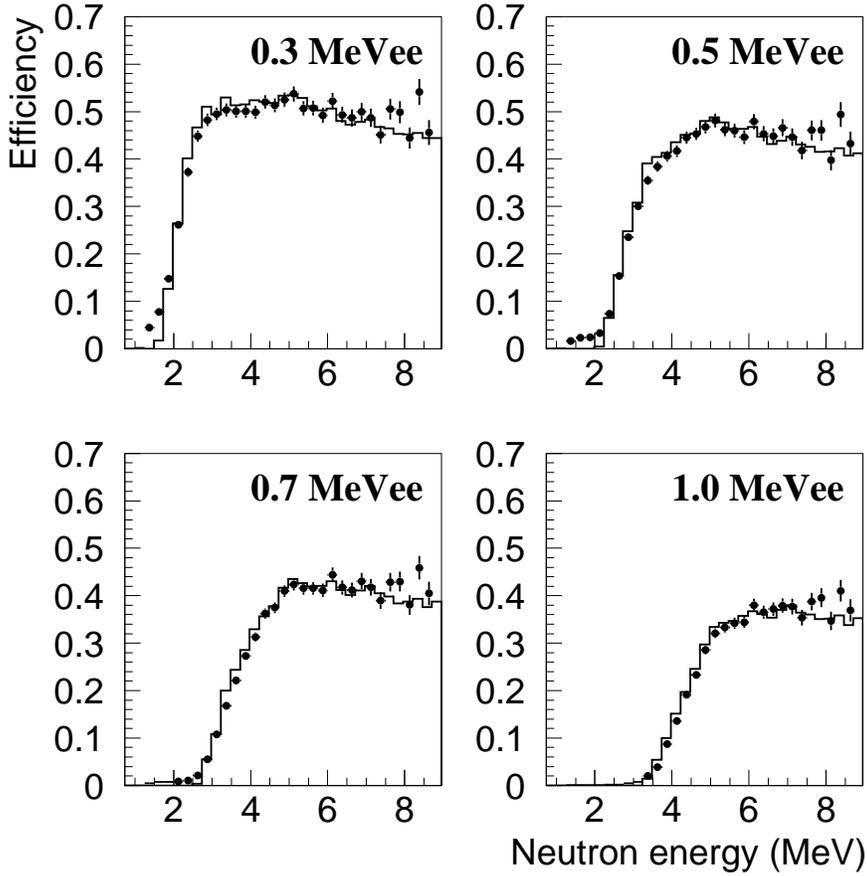}

\caption{\label{cap:Measured-and-simulated}Measured and simulated absolute
detection efficiencies for the Nordball array. Thresholds are 0.3,
0.5, 0.7 and 1.0~MeVee. The full-line histogram represents the MC
simulation. Error bars on the data points denote the statistical uncertainty.}
\end{figure}

The average efficiencies, excluding the bottom detectors of the array,
are presented in Fig. \ref{cap:Measured-and-simulated} and compared
to the MC predictions. \textcolor{black}{At neutron energies very
close to threshold the simulation under predicts the data, especially
for the distributions at 0.3 and 0.5~MeVee threshold. The sharp rise
in the efficiency just above threshold is well described by the MC
simulation, although here the data fall slightly below the model.
Generally, the agreement is best in the region between 5--7 MeV where
the efficiency function flattens out and here the relative difference,
defined as $\left(\varepsilon_{exp}-\varepsilon_{MC}\right)/\varepsilon_{MC}$,
is within 5\%. At higher energies the statistical precision of the
data deteriorates due to the rapid fall in fission yield.} \textcolor{red}{}

\subsection{\label{sub:Deuteron-photodisintegration}The deuteron photodisintegration
measurement}

As a test of the ability of the MC model to correct for systematic
\textcolor{black}{scattering and absorption} effects in neutron photoproduction,
the $\mathrm{^{2}H\left(\gamma,n\right)p}$ cross section was measured
with tagged photons using a $\mathrm{D_{2}O}$ target. The 9~cm long,
cylindrical target had a diameter of 6.6~cm and a significant fraction
of the produced neutrons interacted in the target volume. The Nordball
detectors were positioned as in the $^{252}$Cf efficiency measurement
(Fig. \ref{cap:Nordball-detector-geometry}) with the target centered
at the position previously occupied by the fission source. A 18~mm
thick plastic scintillator placed between target and detectors aided
identification of charged particles, mainly generated by Compton scattering
and pair production in the target. Monoenergetic photons in the range
$E_{\gamma}=14-18$~MeV were produced by the tagged bremsstrahlung
technique using the MAX-lab tagging spectrometer \cite{JOA1}. The
width of a tagged-photon energy bin was 340~keV \textcolor{black}{and
the rate in each plastic-scintillator element of the focal plane was
$\sim400$~kHz.}

The 6--8~MeV neutrons produced by two-body photodisintegration interacted
in the detectors and bulk materials in the target-detector locality.
These interactions were evaluated with the coupled GEANT3-STANTON
simulation, described in Sec. \ref{sub:Geant3-Stanton-code}, to provide
correction factors for the measured neutron yields. 

\begin{figure}[th]
\includegraphics[width=1\columnwidth,keepaspectratio]{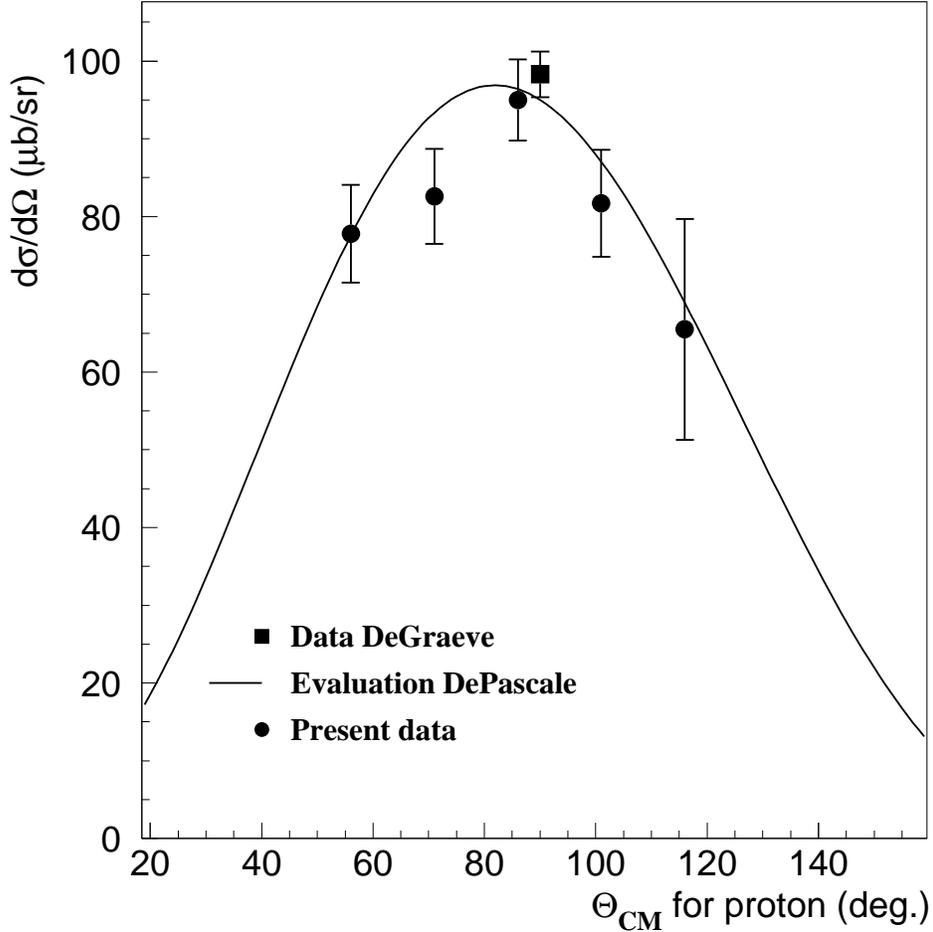}

\caption{\label{cap:sigma_diff}Comparison of the present $^{2}\mathrm{H\left(\gamma,n\right)p}$
differential cross section (solid circles), converted to the equivalent
proton CM angle, for $\mathrm{E}_{\gamma}=14-18$~MeV with the evaluation
of \cite{DePascale} (solid line) and the datum of \cite{DeGraeve}
(solid square). The error bars on the present data denote the statistical
uncertainties while those of \cite{DeGraeve} show the total uncertainty.}
\end{figure}

Fig. \ref{cap:sigma_diff} shows the comparison between the present
data, a high precision measurement at proton $\theta_{\mathrm{CM}}=90^{\circ}$
\cite{DeGraeve}, and a fit to available low energy $\mathrm{^{2}H\left(\gamma,p\right)n}$
data \cite{DePascale}. This fit gives a very good account of the
low energy data set and agrees well with a subsequent low energy evaluation
\cite{Thorlacius}. The standard representation of two-body deuteron
photodisintegration is with respect to the proton CM angle. The good
agreement between data and evaluation supports the ability of the
simulation to take scattering and absorption effects into account,
bearing in mind that $\sim30$\% of the neutrons interacted in the
target volume and also a significant fraction in the plastic-scintillator
sheet. \textcolor{black}{Note that the simple analytical approach
to estimate neutron absorption in each material, on the basis of the
reaction cross section and an exponential attenuation function, underestimates
neutron losses by more than 30\%. This approach fails to account for
elastic scattering effects which are important at low neutron energies. }

The average cross section measured at the 5 angles was $81.4\pm4.5(stat)\pm9.7(sys)\mu\mathrm{b/sr}$
and the parameterisation of \cite{DePascale} gives a value of $84.7\pm1.7\mu\mathrm{b/sr}$
where the uncertainty is due to the uncertainties in the coefficients
of the Legendre expansion of the angular distribution. \textcolor{black}{For
the present data, the sources of systematic uncertainty were detector
solid angle (0.7\%), neutron scattering and absorption (3.2\%), detection
efficiency (5\%), and random-background subtraction (7.6\%). The uncertainty
for scattering and absorption was estimated by comparing neutron attenuation
coefficients from other experiments (Sec. \ref{sub:Other-detectors})
with the predictions of the present MC simulation. }

The present measurement at neutron $\theta_{\mathrm{lab}}=90^{\circ}$
is also consistent with the datum of \cite{DeGraeve} (Fig. \ref{cap:sigma_diff}). 

\subsubsection{\label{sub:Other-detectors}Application to other detector systems}

\textcolor{black}{The present simulation gives an excellent account
of the tagged (by the pion capture reaction $\pi^{-}\mathrm{p}$$\rightarrow$n$\gamma$)
neutron measurement of Ref. \cite{Sawatzky}. This reported not only
the absolute efficiency of a 16 (4x4) cell array of liquid scintillators
(of somewhat smaller volume than the Nordball elements), but also
the magnitude of {}``cross-talk'' effects where neutrons scatter
from one cell to another. The pion capture reaction produced mono-energetic
neutrons of 8.9 MeV kinetic energy which were tagged by detection
of the 129 MeV photon in a large NaI detector. With two-body kinematics
the angular range of the tagged neutron beam was determined by that
of the NaI detector, chosen such that the neutron beam directly illuminated
the four central neutron cells only. The measured detection efficiency
was 21.3$\pm$0.1(stat)$\pm$0.6(sys)\% for the full array while the
simulated efficiency with the present code was 21.6$\pm$0.1(stat)\%.
Magnitudes and shapes of the pulse-height distributions for cross-talk
events were also well reproduced. Further comparisons of simulation
predictions with neutron measurements} \cite{Desesquelles,von-Edel}
also show good agreement.

\section{\label{sec:Conclusions}Conclusions}

The scintillation response and detection efficiency of the Nordball
liquid scintillator array for neutrons was measured using the well
known fission, fast-neutron yield from a $^{252}$Cf source. Using
the measured recoil proton pulse-height response as input to a MC
computer model of Nordball, good agreement between simulated and measured
detection efficiencies was achieved for neutron energies up to 8~MeV.
Above this the fission yield is too low for statistically significant
calibrations. 

The MC simulation is intended to provide systematic correction factors
for neutron photoproduction cross sections and this was checked by
measuring the well known $^{2}$H$\left(\gamma,\mathrm{n}\right)\mathrm{p}$
cross section in the energy range $E_{\gamma}=14-18$~MeV. Comparison
with parameterisations of the $\gamma+\mathrm{d\rightarrow\mathrm{p+n}}$
differential cross section and previous data suggest that the uncertainty
of the MC calculation is within 5\%.

In the future we plan to convert of the present GEANT-3 based code
to the more modern object-oriented GEANT-4 framework. This is now
used extensively to model ionising-radiation interactions in matter
at all energies and has branched into many applications including
medical physics.

\subsection*{Acknowledgments}

The authors wish to thank the staff of MAX-lab for their excellent
provision of the electron beam and general laboratory infrastructure.
We thank the following for their financial support: the UK Engineering
and Physical Sciences Research Council, the Swedish Natural Sciences
Research Council and the Italian Instituto Nazionale di Fisica Nucleare.
One of the authors (A.R.) expresses thanks to Deutscher Akademischer
Austausch Dienst (DAAD) for their support.

\end{document}